\documentclass[12pt]{spie} 

\usepackage{cite}
\usepackage{graphicx}
\usepackage{caption}
\usepackage{subcaption}
\usepackage{amsmath}
\usepackage{float}

\title{ Spatial extension of excitons in triphenylene based
  polymers given by range-separated functionals}
\author{B. Kociper and
  T.A. Niehaus\footnote{Corresponding author: thomas.niehaus@physik.uni-r.de}
\skiplinehalf
Department of Theoretical Physics, University of Regensburg,
93040 Regensburg, Germany}
\begin{document}
\maketitle
\begin{abstract}

Motivated by an experiment in which the singlet-triplet gap in triphenylene based
copolymers was effectively tuned, we used time dependent density functional theory (TDDFT) to reproduce the main results. By means of
conventional and long-range corrected exchange correlation
functionals, the luminescence energies and the exciton localization were calculated
for a triphenylene homopolymer and several different copolymers.  The
phosphorescence energy of the pure triphenylene chain is predicted
accurately by means of the optimally tuned long-range corrected
LC-PBE functional and slightly less accurate by the global
hybrid B3LYP. However, the experimentally observed fixed
phosphorescence energy could not be reproduced because the localization
pattern is different to the expectations: Instead of localizing on the
triphenylene moiety - which is present in all types of
polymers - the triplet state localizes  on
the different bridging units in the TDDFT calculations. This leads to different triplet emission
energies for each type of polymer. Yet, there are clear indications that
long-range corrected TDDFT has the potential to predict the triplet
emission energies as well as the localization behavior more accurate
than conventional local or semi-local functionals.

\end{abstract}

\section{\label{sec:intro}Introduction}

Today, conjugated polymers play an important role in applications like
organic light emitting diodes (OLEDs) \cite{Grimsdale2009} or in
organic photovoltaics.\cite{Brabec2003,Gunes2007} A key quantity for
optimal device performance is the singlet-triplet gap ($\Delta E_{ST}$), because a
small gap favours a higher emitter efficiency.
\cite{Medina2012} Usually electrical injection of charges into an OLED
leads to the generation of singlet and triplet excitations, where the decay of the latter
is mostly non-radiative. To promote radiative triplet recombination,
heavy-atom centers are employed to promote spin-orbit
coupling, leading to a higher phosphorescence
probability.\cite{Baldo1998,Cleave1999,Yang2006,Ying2009,Haneder2008} An
exception are triphenylene and some other polycyclic aromatic
hydrocarbons: The phosphorescence from the triplet state is observed at
low temperatures without incorporation of heavy atoms.
\cite{Clark1969,Kellogg1964,Langelaar1968} A small $\Delta E_{ST}$ does
not only increase the rate for triplet emission directly,\cite{Haneder2008} it allows also   
for thermal repopulation of the lowest singlet state (S$_1$), and a
following delayed fluorescence. This {\em singlet
harvesting} approach was recently put forward by Yersin and co-workers
to enable the collection of both singlet and triplet excitons that
are created by charge injection in OLED.\cite{Czerwieniec2011,Yersin2011}

\begin{figure}[H] 
\centering
\includegraphics[width=0.55\textwidth]{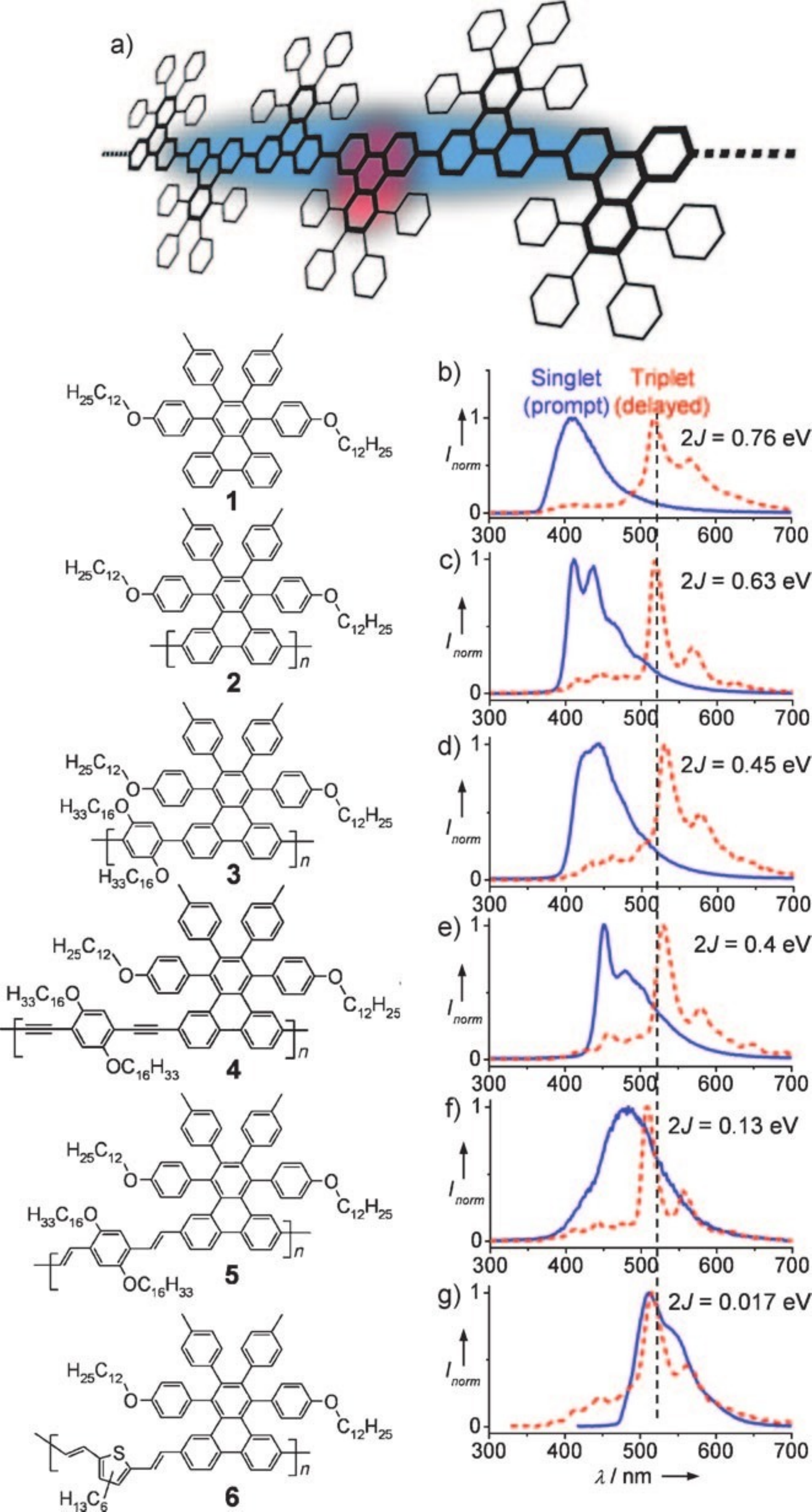}
\caption{Fluorescence (prompt emission) and phosphorescence (delayed
  emission) from a triphenylene-based monomer and conjugated
  copolymers, dispersed in a polystyrene matrix at 25 K. a)
  Delocalization of singlet excitations (blue) with triplets (red)
  localized at the triphenylene unit. b) - g) Singlet (solid blue
  line, integrated 0-2 ns after excitation) and triplet (dashed red
  line) spectra of the monomer \textbf{1} (0.1 - 1.1 ms delay after
  excitation), the homopolymer  \textbf{2} (9 - 10 ms delay), the
  para-phenylene copolymer  \textbf{3} ($1Ð2$ ms delay), the
  ethynylene copolymer  \textbf{4} (1.5 - 2.5 ms delay), phenylene
  vinylene copolymer  \textbf{5} (0.02 - 1.02 ms delay), and the
  thienylene vinylene copolymer  \textbf{6} (0.05 - 5.05 ms
  delay). The exchange splitting $2J$ is estimated from the peak
  separations. The dashed black line indicates the average triplet
  peak position. Reprinted with permission from Ref.~[\citeonline{Chaudhuri2010}].}
\label{fig:fig_1}
\end{figure}

In a recent experiment,\cite{Chaudhuri2010} the group of Lupton
investigated the singlet-triplet splitting in a series of
triphenylene based  phosphorescent copolymers and showed that $\Delta
E_{ST}$ may be effectively tuned (see Figure \ref{fig:fig_1}). The
copolymers exhibit all a similar triplet spectrum at the same energetic
position,  whereas the energy of the singlet peak strongly
depends on the details of the polymer backbone. The fact that the
singlet emission shifts towards the triplet emission from monomer
\textbf{1} to polymer \textbf{6}, whereas the triplet level remains
almost unchanged, implies that singlet and triplet excitations can
form on different parts of the conjugated system.\cite{Glusac2007}
This indicates that the splitting can be tuned and that eventually even the regular level ordering of singlet and triplet may be inverted.\cite{Chaudhuri2010}  

Our goal in this study is to investigate whether time dependent
density functional theory (TDDFT) with state of the art exchange
correlation (xc) functionals is capable of describing this highly unusual
photo-chemical system. The interest arises, because popular local and
semi-local xc functionals that are widely used for condensed matter
applications fail to provide a reasonable description of the electronic
structure in extended $\pi$-conjugated systems. Typical shortcomings
for polymers are an underestimated bond length alternation\cite{Jacquemin2010b,Korzdorfer2012}, an overestimation of
electronic polarizabilities\cite{Champagne2000} and conductances\cite{Ke2007}, the instability of
polaronic excitations\cite{Niehaus2004}, and an underestimation of the
optical gap. Many of these deficiencies have been related to the
incomplete cancellation of self-interaction effects due to approximate
exchange functionals which lead to an overly delocalized electron
density.\cite{Cohen2008} To a large extent these difficulties may be
overcome by range-separated xc functionals, that recently achieved a
lot of attention. The main idea is to separate the electron-electron
interaction into a short-range domain, where conventional semi-local
functionals work well, and a long-range domain, where non-local
Hartree-Fock (HF) exchange reduces self-interaction errors
significantly.\cite{Mori-Sanchez2006,Haunschild2010} In the early
formulations,\cite{gill96,Savin1996} the domains were defined by a
single empirical parameter valid for all systems. Following the work of Baer and coworkers
\cite{Baer2010} this parameter may also be tuned
on a system per system basis to match known conditions for the exact
functional (e.g., the ionization potential being equal to the highest
occupied orbital energy), escaping the need for an empirical fit to a
large training set of molecules.  Earlier studies indicate the importance of non-local exchange
(either within conventional global hybrids or tuned range-separated xc
functionals) also for the correct estimation of
exciton sizes.\cite{Tretiak2005,Nayyar2011,Pandey2012} As the mentioned
experiment focuses on luminescence properties of the triphenylene
polymers, we are especially interested in the spatial extension of the
lowest singlet and triplet states after excited state relaxation has
taken place.     

\section{\label{sec:computational} Computational Details}
\begin{figure}[t] 
\centering
\includegraphics[width=0.45\textwidth]{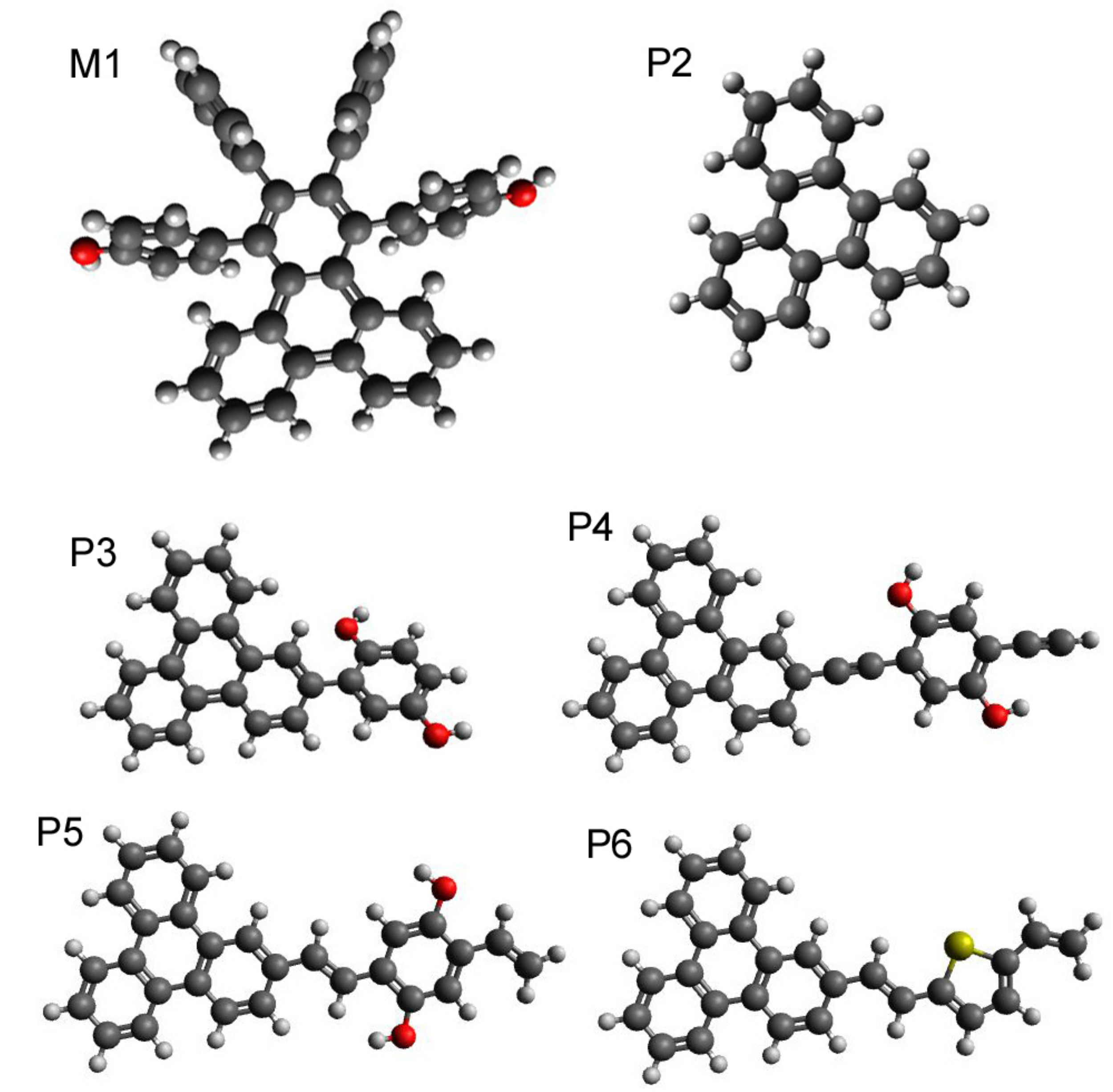}
\caption{ Model
  structure for the triphenylene based monomer {\bf M1} and basic repeat units of the polymers  {\bf P2}-{\bf P6}  used in
  this study. }
\label{fig:poly}
\end{figure}

In order to build realistic but still numerically tractable models for
the copolymers depicted in Figure \ref{fig:fig_1}, we removed the
toluene and phenol groups bonded to the triphenylene unit and all alkane chains to
arrive at the polymer building blocks depicted in
Figure \ref{fig:poly}. Mass spectra from the experiment yield the
number of basic units for each polymer. Approximately 48-52 units
formed the triphenylene polymer \textbf{2}, around 32-34 units the
polymer \textbf{3}, 15-17 units the polymer \textbf{4} and only 5-7
units the polymer \textbf{5} and 4-6 units the polymer \textbf{6}.  As
the TDDFT calculations for chains with more than 260 atoms were
computationally too demanding, the number of basic units in a chain
was limited to $n$=6. Therefore the TDDFT results for polymer \textbf{5}
and \textbf{6} can be directly compared to experiment, whereas the other polymer types \textbf{2},
\textbf{3}, and \textbf{4} could not be calculated up to comparable length
scales. In the following, oligomers with $n$ repetitions of the basic
unit $\alpha$ are denoted as {\bf P}${\bf \alpha}_n$. We also considered the monomer
\textbf{1}, which differs from the corresponding structure in the
experiment only in the missing alkane chains and methyl groups, which
should have a marginal influence on the optical properties. All
structures were optimized without symmetry constraints using the
hybrid B3LYP\cite{B3}  functional  and a
6-31G* basis set with the {\tt NWChem 6.0}
code. \cite{Valiev2010} In contrast to local and semi-local functionals, B3LYP provides
more realistic
values for the bond length alternation in conjugated polymers, especially for short 
chains.\cite{Jacquemin2010b}

For the excited state calculations we compare the semi-local PBE
functional,\cite{perdew1996gga} the hybrid B3LYP, and the
range-separated LC-PBE functional.    
Range-separated functionals partition the Coulomb operator into a
short-range and long-range component, ruled by the parameter $\gamma$,
for example by utilizing the standard error-function $\text{erf}(x)$ \cite{gill96,Savin1996}:
\begin{equation}  
\label{partition}
\frac{1}{r}=\frac{1-\text{erf}(\gamma r)}{r}+\frac{\text{erf}(\gamma r)}{r}.
\end{equation}
The first term is a Coulomb operator decaying to zero on a length
scale of $\approx 1/\gamma$ and is therefore short-ranged (SR), while
the second term dominates at large $r$ accounting for the long-range
(LR) behavior. Both conventional pure and hybrid functionals may be
extended according to this idea. In the former case the functional
decomposition reads
\begin{equation}
  \label{lc-pbe}
  E_\text{xc} = E_\text{x,DFT}^\text{SR}(\gamma) +
  E_\text{x,HF}^\text{LR}(\gamma)  +  E_\text{c,DFT}.
\end{equation}
In the case of LC-PBE, $E_\text{x,DFT}^\text{SR}$ is the short-range
form of the gradient-corrected PBE exchange functional\cite{Yanai2004},
$E_\text{x,HF}^\text{LR}$ denotes the Hartree-Fock exchange energy
evaluated with the long-range part of the interaction in Equation
\ref{partition}, while the correlation part $E_\text{c,DFT}$ is left
unchanged with respect to the usual form of PBE. Instead of working with a
fixed value of the range-separation parameter $\gamma$, we follow the
ideas of Baer and co-workers\cite{Baer2010} to determine its value for each system
separately. To this end, one considers the following error function 
\begin{equation}
\label{iptune}
\Delta_\text{IP}(\gamma)=\left|\epsilon^{\gamma}_\text{HOMO} - \left[E_\text{tot}^{\gamma}(N)-E^{\gamma}_\text{tot}(N-1)\right]\right|,
\end{equation}
where $\epsilon^{\gamma}_\text{HOMO}$ denotes the orbital energy of
the highest occupied molecular orbital (HOMO) and
$E_\text{tot}^{\gamma}(N)$ and $E^{\gamma}_\text{tot}(N-1)$ stand for
the total energy of the neutral system with $N$ electrons and the ionized one. The latter is
obtained by means of a spin unrestricted calculation. Minimization of
$\Delta_\text{IP}$ yields the optimal range-separation parameter and
enforces a condition that the exact exchange-correlation functional
should exhibit, namely that the negative of the HOMO energy equals the
ionization potential. Key to the success of this scheme is the fact
that ionization energies obtained from total energy differences are
typically quite accurate in DFT.   

For the different functionals, vertical excitation energies were
obtained from frequency domain TD-DFT in linear response
(a.k.a.~RPA or Casida approach)\cite{Casida1995}.  The Tamm Dancoff
approximation \cite{Hirata1999} (TDA) was employed, because there is
evidence that this improves the lowest lying singlet and triplet states
as explained in the next section. Additionally, the use of the TDA enabled
the calculation of longer polymer chains, as the computational effort
is reduced. The 6-31G* basis set was used for all excited state
computations. Test calculations on the basic triphenylene unit
({\bf P2}$_1$) with a basis set of triple-$\zeta$ quality and polarization
functions on all atoms led to a change of only 0.02 eV for $S_1$ and
$T_1$ excitation energies. This deviation is expected to be much smaller than the
error introduced by the molecular models and the available approximate
exchange-correlation functionals. 

Fluorescence and phosphorescence  energies were obtained by optimizing the geometry on the
$S_1$ and $T_1$ potential energy surface, respectively. At the resulting
geometry, vertical emission energies were computed by means of linear
response TD-DFT to estimate the
energy of maximum emission. Because of the high computational cost,
excited state optimization in the singlet state could only be
performed for the polymers {\bf P2} and {\bf P6}. The {\tt Amsterdam Density
Functional} program (ADF) \cite{TeVelde2001, Guerra1998, ADF2012} with
DZP basis set was used
for this purpose. Optimization in the $T_1$ was carried out again with
{\tt NWChem 6.0} using ground state DFT with a spin multiplicity of
three and 6-31G* basis.  The geometry of polymer \textbf{4} was
difficult to converge especially in the triplet state. The
corresponding results are incomplete and therefore left out in the following discussion.                

\section{\label{sec:result}Results}
\subsection{\label{sec:gsgeo} Polymer structure}  

While the polymers were dispersed and  embedded  in a polystyrene matrix in the experiment, our calculations correspond
to single molecules in the gas phase. It is difficult to quantify the
influence of the surrounding medium on the polymer structure
precisely, but it seems natural to assume that a planarization of the
polymer backbone takes place to reduce steric hindrance. Nevertheless,
we do not enforce perfect planarity in the optimizations, since recent
thin film results indicate inter- and intra-segment twist in
a variety of copolymers.\cite{Donley2005,Lupton2010,Habuchi2011} Without symmetry
constraints, twisted geometries are obtained in the ground state for polymer \textbf{2} and
\textbf{3}, whereas for polymer  \textbf{5} and  \textbf{6} planar
geometries showed to be energetically favored. For
\textbf{P2} we additionally tested the impact of forced planar
structures on the results. The $T_1$ absorption and emission energies for the planar
compounds lie in average 0.23 eV and 0.13 eV, respectively,  below the ones
for twisted geometries. Important for the following is that both type
of structures exhibited very similar localization characteristics of
the excited state. 
\begin{table}[t]
  \centering
    \begin{tabular}{llccccccc}
    && \multicolumn{3}{c}{S$_1$ absorption} &
      \multicolumn{2}{c}{T$_1$ absorption} & \multicolumn{2}{c}{T$_1$ emission}\\
      \cline{3-5}  \cline{6-7}  \cline{8-9}
& Functional  & Energy & f  &  $\Phi_{i}^{a}$  & Energy  &  $\Phi_{i}^{a}$ &
Energy  & $\Phi_{i}^{a}$\\\hline
{\bf M1} & PBE & 3.18 & 0.07 &  $\Phi_{H}^{L}$   & 2.84 & $\Phi_{H}^{L}$  & 1.95 &$\Phi_{H}^{L}$ \\
& B3LYP & 3.76 & 0.01 & $\Phi_H^{L+1}, \Phi_{H-1}^{L}$     & 3.02 &$\Phi_{H}^{L}$ &
2.05 & $\Phi_{H}^{L}$\\
& LC-PBE & 3.88 & 0.01 &$\Phi_H^{L+1}, \Phi_{H-1}^{L}$ & 3.14 & $\Phi_{H}^{L}$&
2.06 & $\Phi_{H}^{L}$ \\\\
{\bf P2}$_1$ &  PBE & 3.68  & 0.00 & $\Phi_H^{L+1}, \Phi_{H-1}^{L}$  &
3.17 & $\Phi_H^{L}, \Phi_{H-1}^{L+1}$  & 2.64  & $\Phi_{H}^{L}$ \\
&  B3LYP & 4.03 & 0.00 & $\Phi_H^{L+1}, \Phi_{H-1}^{L}$ & 3.21 & $\Phi_H^{L},
\Phi_{H-1}^{L+1}$ & 2.63 & $\Phi_{H}^{L}$ \\
& LC-PBE & 4.29 & 0.00 & $\Phi_H^{L+1}, \Phi_{H-1}^{L}$ &   3.36 & $\Phi_H^{L},
\Phi_{H-1}^{L+1}$ & 2.77 & $\Phi_{H}^{L}$ \\[0.2cm]
    \end{tabular}
  \caption{S$_1$/T$_1$ absorption and T$_1$ emission energies
    (in eV) for the monomer {\bf M1} and {\bf P2}$_1$ (i.e.~Triphenylene). Oscillator strengths (f) are given for
    singlet-singlet transitions only. Singly excited
    determinants 
    that contribute significantly to the excited state wave function
    are denoted as $\Phi_i^a$ (for a transition from orbital $i$ to $a$).
    The highest occupied molecular orbital (HOMO) is abbreviated as
    $H$, the lowest unoccupied one (LUMO) as $L$. The range-separation
    parameter was determined to be $\gamma$ = 0.17 a$_0^{-1}$
    for {\bf M1} and $\gamma$ = 0.23 a$_0^{-1}$ for {\bf P2}$_1$. The
    experimental T$_1$ emission energy for {\bf M1} is 2.37 eV \cite{Chaudhuri2010}, and
    the S$_1$ absorption energy for triphenylene in solution was
    reported to be 3.78 eV.\cite{Levell2010}}
  \label{tab:monomer_tab}%
\end{table}%

\begin{figure}[t]
\centering
  \begin{minipage}[w]{100pt}
    \includegraphics[width=100pt]{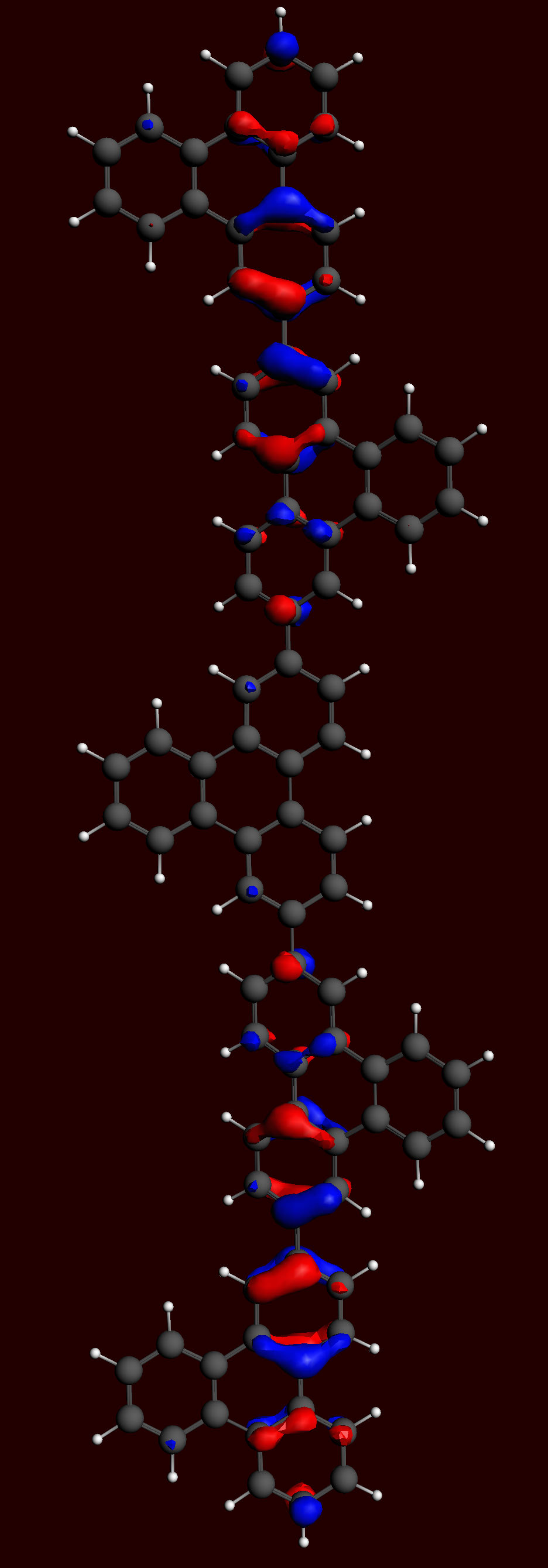}   
  \end{minipage} 
  \begin{minipage}[w]{100pt}
    \includegraphics[width=100pt]{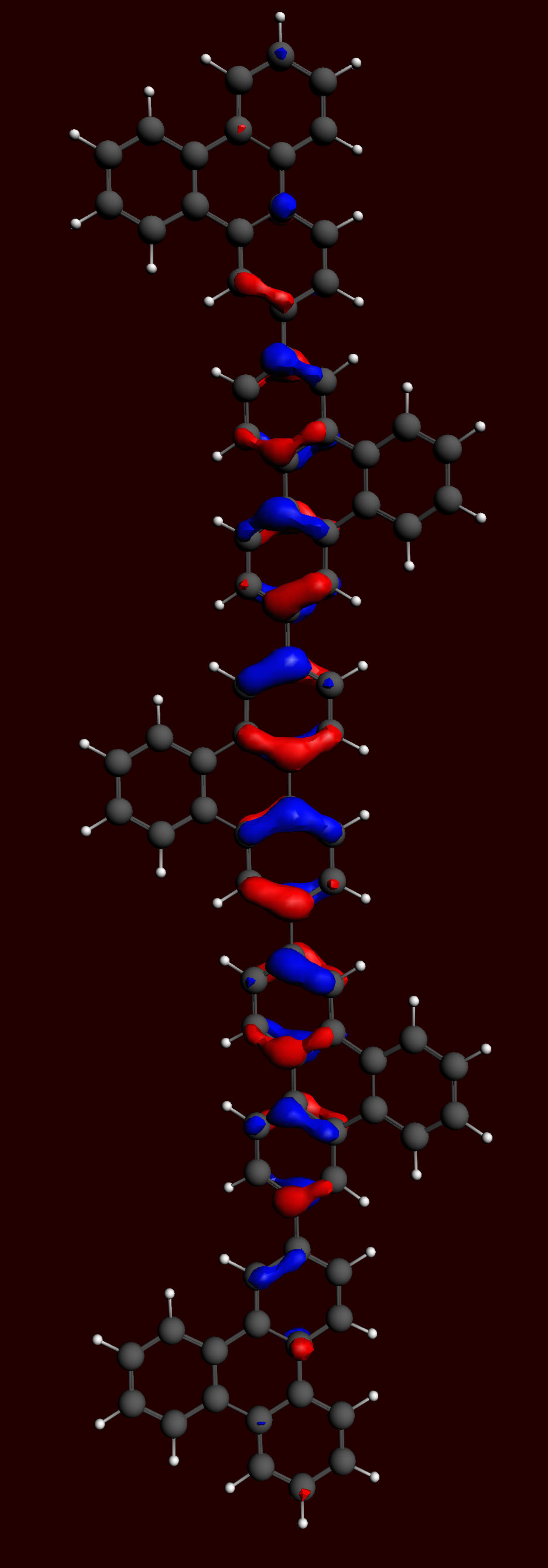}  
  \end{minipage}
   \begin{minipage}[w]{100pt}
    \includegraphics[width =100pt]{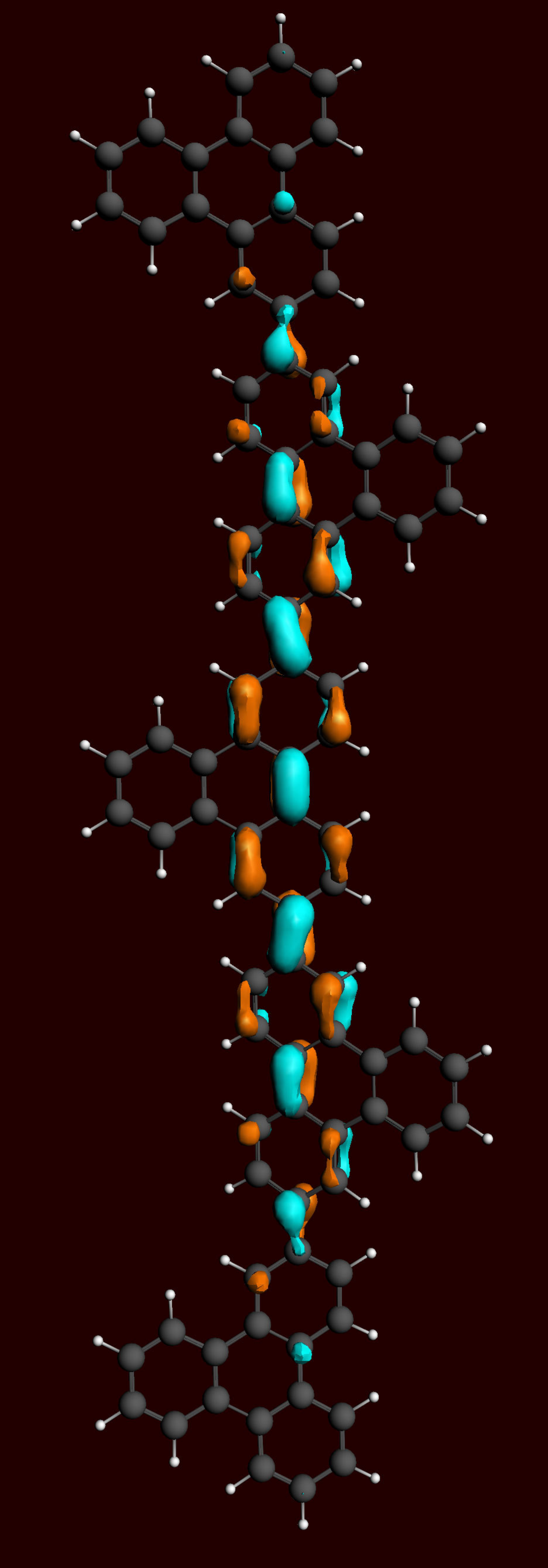}   
  \end{minipage} 
  \begin{minipage}[w]{100pt}
    \includegraphics[width =100pt]{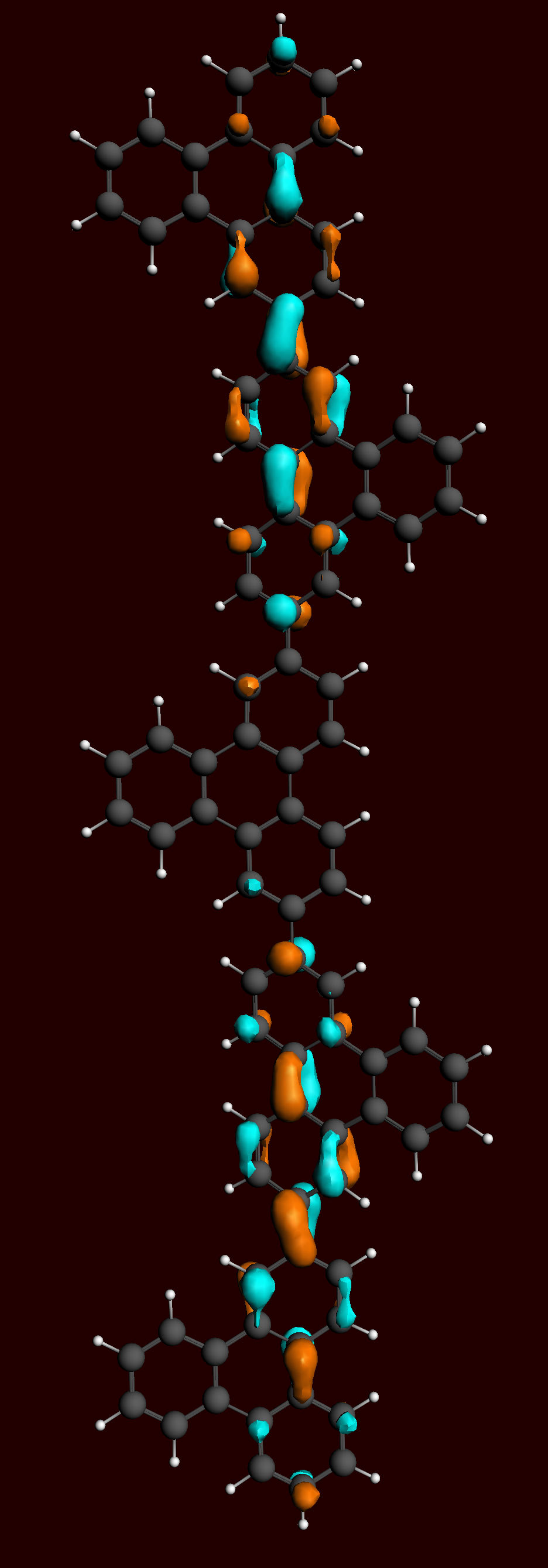}  
  \end{minipage}
  \caption{From left to right: molecular orbitals HOMO-1, HOMO, LUMO and LUMO+1
    of {\bf P2}$_5$ at the
    B3LYP/DZP level of theory.}
    \label{fig:edens}
\end{figure}

In order to assess the significance of our reduced molecular models,
the absorption and luminescence energies of {\bf M1} and {\bf
  P2}$_1$ are compared in Table \ref{tab:monomer_tab}. The two
compounds differ only in the attached functional groups, which are
left out in the longer polymer models. It turns out, that for these
monomers the influence of the aromatic substituents is sizable and
leads to a red shift (depending on the functional) of 0.2-0.4 eV  in
absorption and of 0.6 eV in emission. Interestingly, unconstrained
optimization of {\bf M1} in the T$_1$ state led to rotation of the
attached benzene moieties and a concomitant closing of the optical
gap. Such a volume demanding structural change is not likely 
in a matrix environment. Therefore, the dihedral angle of the functional
groups was fixed with respect to the triphenylene plane and the
resulting phosphorescence energies in Table \ref{tab:monomer_tab} are
obtained. All functionals show a reasonable agreement with the
experimental result of 2.37 eV given by Lupton.  For triphenylene, the
S$_0\to$S$_1$ transition is symmetry forbidden, and is allowed but
very weak for {\bf M1}. This indicates that the additional functional
groups perturb and influence the electronic structure, but do not
completely change it. Further evidence for this fact is given by the
spatial extension of the frontier molecular orbitals for longer
oligomers depicted in Figure \ref{fig:edens}. These are  located
  mainly on the inner triphenylene rings along the chain
  direction.  Compared to the non-neglible effect for the monomer, additionally attached benzene units
  may hence play a less important role in the polymer
  limit. It should also be mentioned at this point, that the frontier
  orbitals are delocalized over the full chain. The excited state
  localization schematically depicted in Figure \ref{fig:fig_1}a is
  not present in absorption.

\subsection{\label{sec:order} Ordering of the two lowest singlet
  excited states}
Special attention was paid to the two lowest singlet excited
states. One of these, which is commonly denoted with $L_a$ is
mainly a HOMO $\rightarrow$ LUMO excitation. The
other one, $L_b$, consists by more or less equal parts of the
HOMO-1$\rightarrow$ LUMO and HOMO $\rightarrow$ LUMO+1
transition. For linear acenes as well as triphenylene and
also some nonlinear PAHs (polycyclic aromatic hydrocarbons), the $L_a$ is
stated to have  charge-transfer (CT) character, whereas the $L_b$ is
covalent like the ground state.\cite{grimme2003set,Wang2007b} To find out whether one of
the first excited states is charge-transfer like is important, because
it is well known,\cite{dreuw2003lrc,dreuw2004ftd,Wanko2004} that TDDFT predicts those
energies with large errors increasing with system size.
\cite{Pogantsch2002} Several studies have shown,\cite{Peach2008,Wong2010,Richard2011}
that range-separated functionals
decrease this error for CT states; in some cases even its
system size dependence.\cite{Richard2011} Some care is necessary 
in the evaluation of triplet states as they can be sensitive to
instabilities in the ground state \cite{Jacquemin2010c} - but this may be
cured by a system dependent determination of the range separation parameter $\gamma$ plus the TDA,
\cite{Sears2011} which is also favorable because it reduces memory requirements. \\ The calculations of all polymers show that the
investigated chains generally do not show a wrong ordering of
the two lowest singlet  excited states. Only one molecule, {\bf P2}$_3$, switches its lowest lying
singlet states when using the LC-PBE functional instead of B3LYP or
PBE. Since fluorescence is strong in the experiment, one expects the
S$_1$ to be strongly dipole allowed. This is also found in the calculations,
where (apart from the shorter {\bf P2}$_2$ and {\bf P3}$_1$) the S$_1$ is  dominated by the HOMO $\rightarrow$ LUMO
transition and exhibits a sizable oscillator strength that increases
linearly with chain length for the B3LYP and LC-PBE functionals. 
 
\subsection{\label{sec:lrc_scaling}Scaling of the range separation
  parameter}
\begin{figure}[t] \centering
\includegraphics[width=0.50\textwidth]{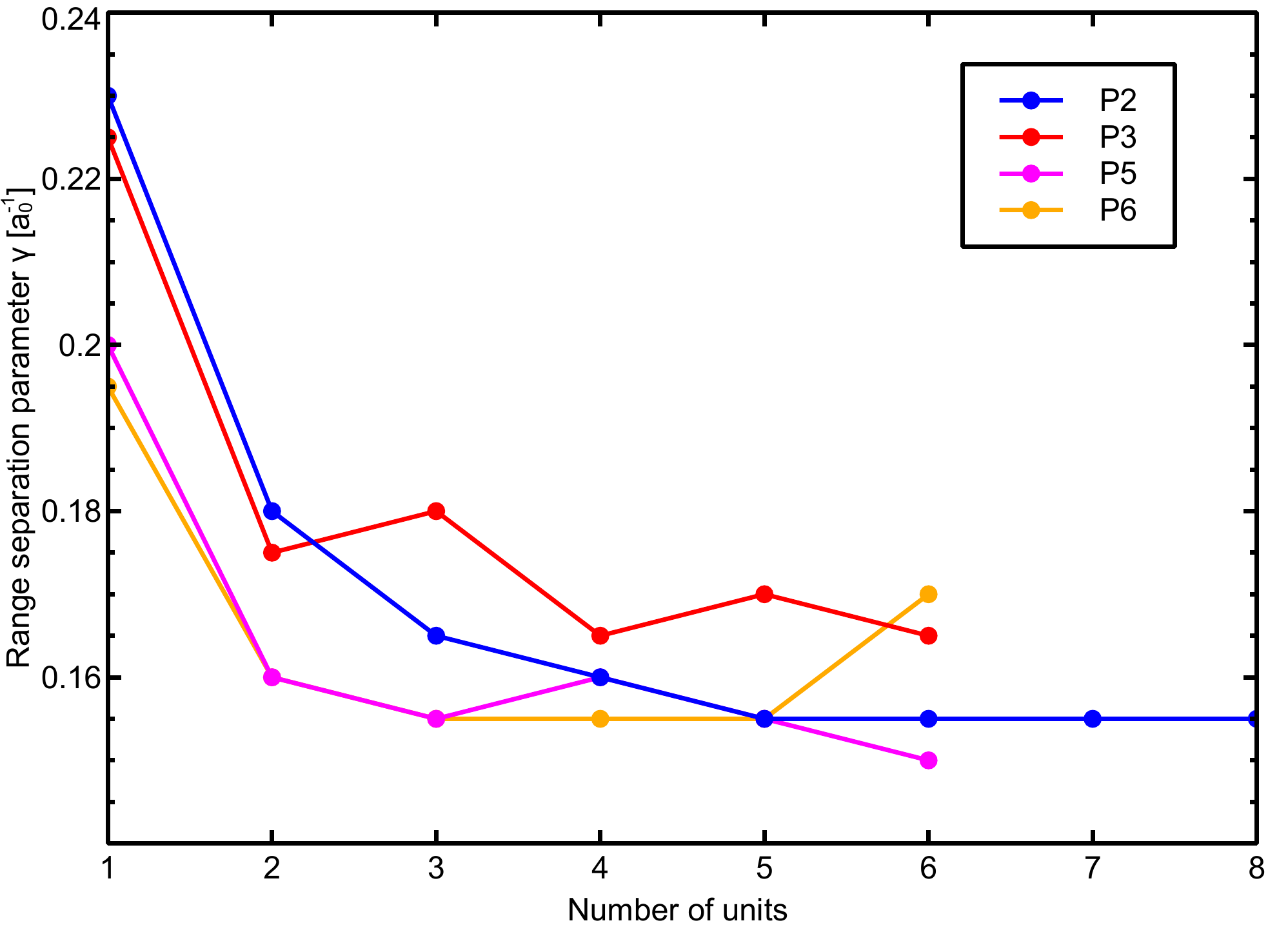}
\caption{Optimal range separation parameter in units of the inverse
  Bohr radius (a$_0$) versus number of polymer units. The minimum of
  $\Delta_\text{IP}$ was found by varying $\gamma$ in increments of 0.005 a$_0^{-1}$.  }
\label{fig:lrc_scaling}
\end{figure}
If obtained from a tuning criterion like Equation \ref{iptune}, the parameter $\gamma$ determines the optimal range of a local
description of exchange effects by means of conventional Kohn-Sham
DFT. It should therefore depend on the actual electronic structure of the
system of interest, which is encoded in system specific variables like
the electron density or microscopic dielectric function. A recent
investigation indicates a close 
correlation of the range separation parameter with the
conjugation length in polymers.\cite{Korzdorfer2011,Pandey2012} For alkanes and
polythiophenes are clear saturation of $1/\gamma$ with system size was
observed, in full agreement with a finite conjugation length. In
contrast, this parameter increased linearly for oligoacenes up to nine
benzene rings.\cite{Korzdorfer2011} The results depicted in Figure \ref{fig:lrc_scaling}
confirm a saturation of $\gamma$, which is most clearly seen for {\bf P2}. The
range separation parameter depends strongly on the chemical structure of the
copolymer and seems to converge for n $>$ 4 to values between 0.15
a$_0^{-1}$ and 0.17 a$_0^{-1}$. The optimal range separation parameter
therefore decreases with system size, but it does not tend to
zero. This means that the tuned LC-PBE functional differs from PBE also
in the polymer limit.

\subsection{\label{sec:energies}Transition energies}
\begin{figure}[t]
        \centering
        \begin{subfigure}[t]{0.5\textwidth}
                \centering
                \includegraphics[width=\textwidth]{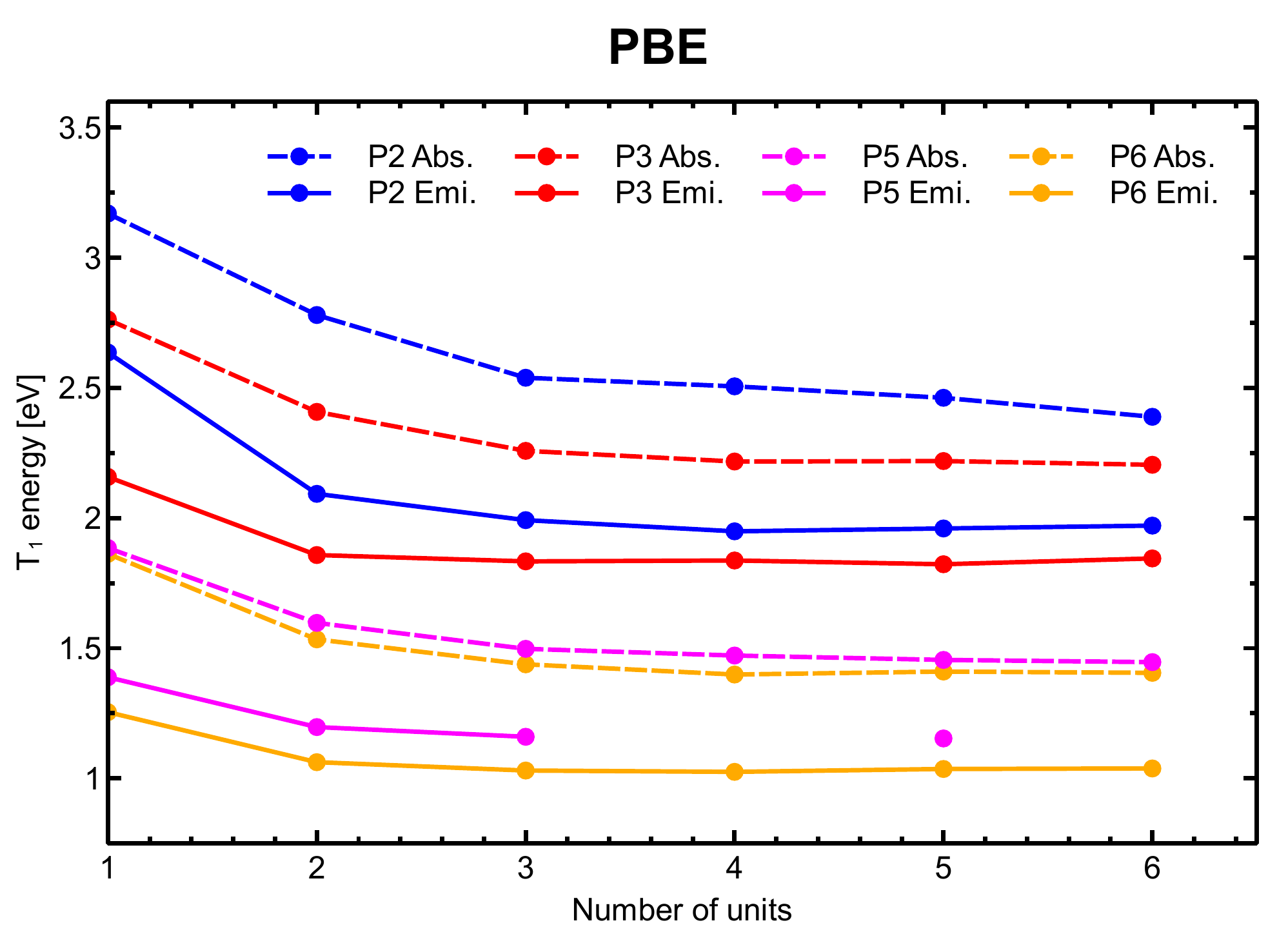}
        \end{subfigure}%
        \begin{subfigure}[t]{0.5\textwidth}
                \centering
                \includegraphics[width=\textwidth]{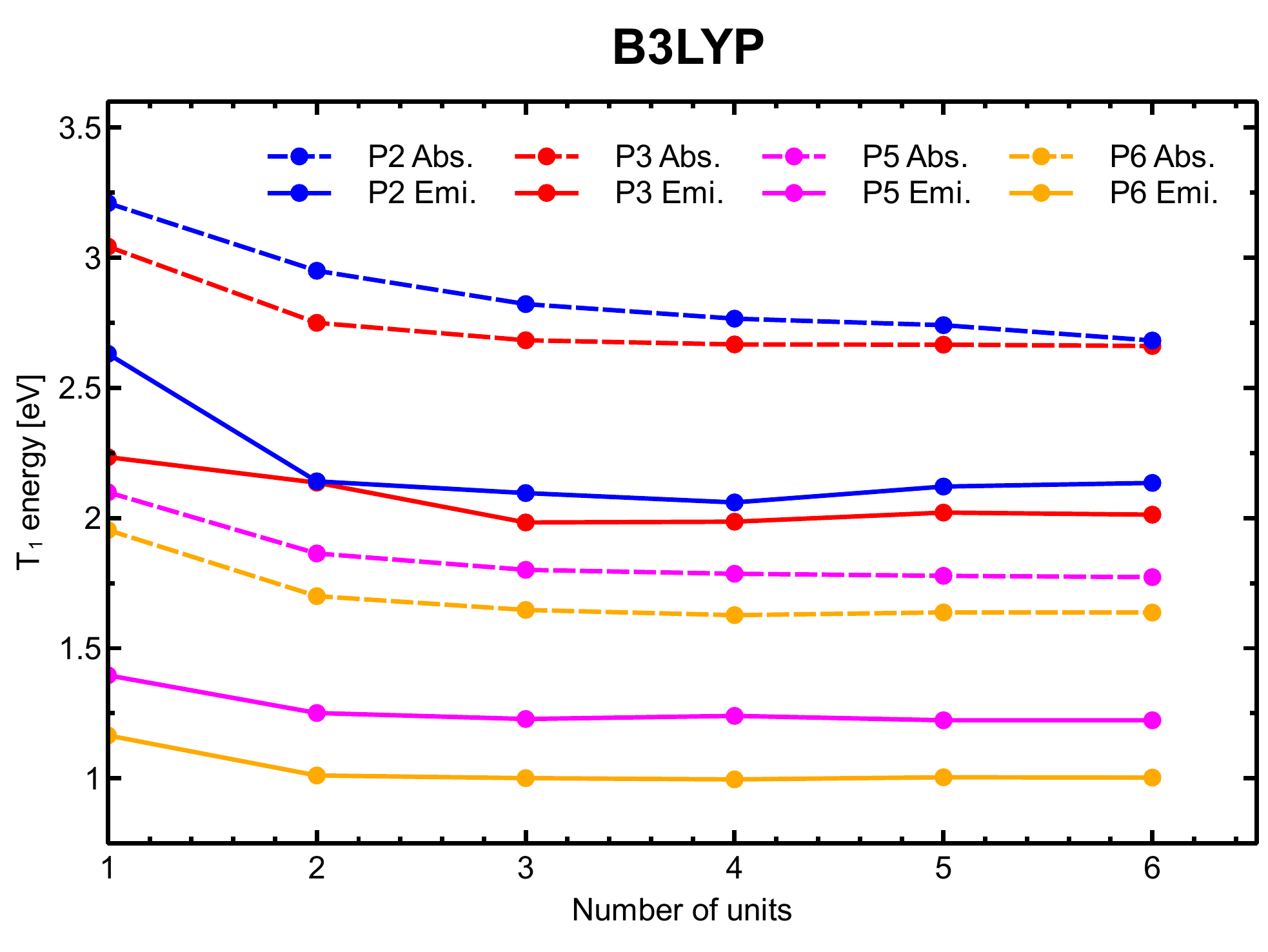}
        \end{subfigure}
        \begin{subfigure}[t]{0.5\textwidth}
                \centering
                \includegraphics[width=\textwidth]{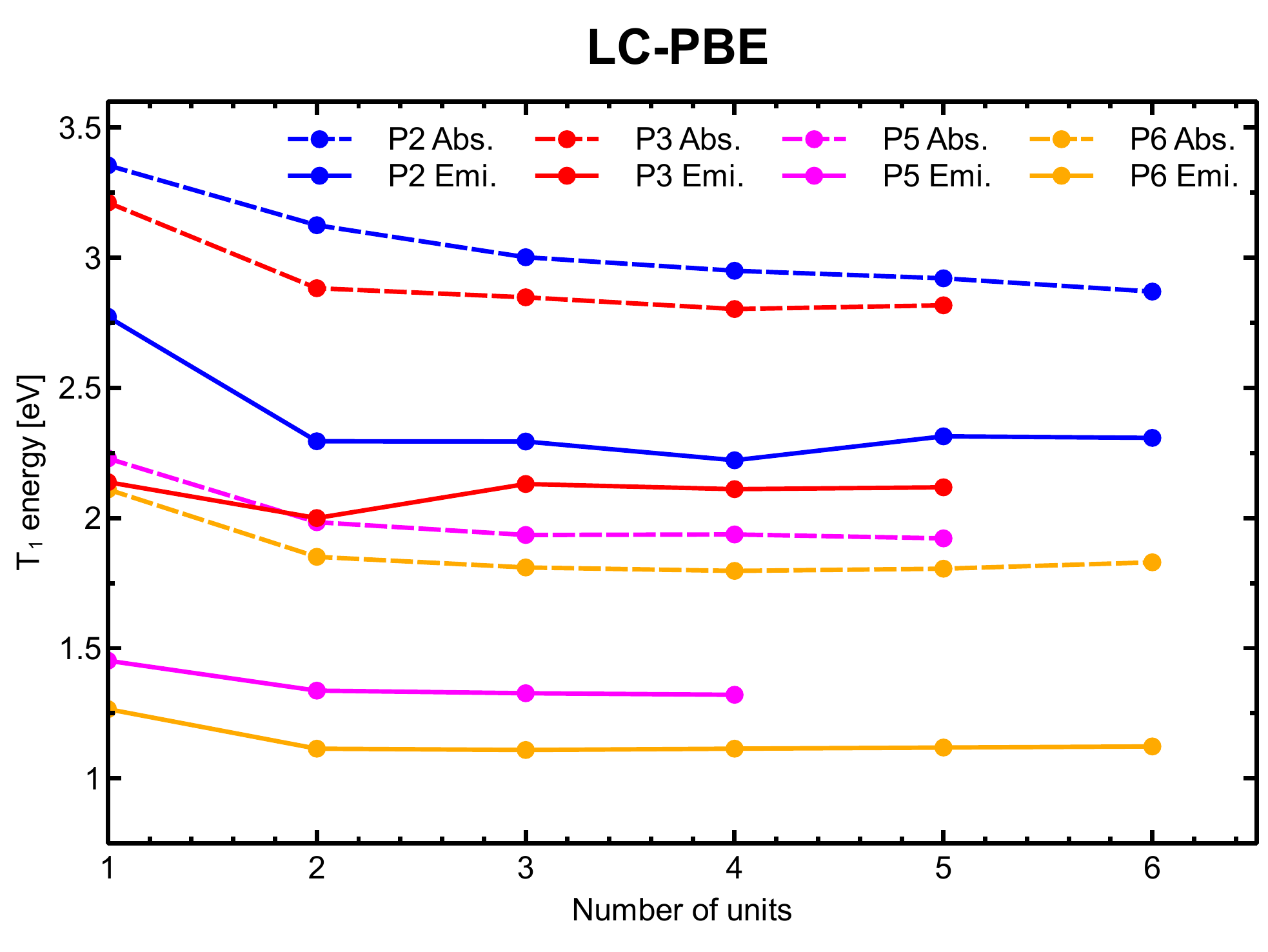}
        \end{subfigure}
\caption{$T_1$ absorption and emission energies with increasing polymer length
at the PBE, B3LYP and LC-PBE level. Missing datapoints correspond to
unconverged calculations. The triplet emission energy depends on
the type of polymer, whereas experimentally \cite{Chaudhuri2010} the
triplet phosphorescence is measured at an almost fixed value of 2.37
eV for \textit{each} copolymer.}
\label{fig:all_polys}       
\end{figure}

We now move to central part of the present investigation, the electronic
excited states. Figure \ref{fig:all_polys} shows the triplet
  absorption and emission energies as a function of the oligomer
  length. As expected from a simple particle-in-a-box model, the
  excited state energies decrease with increasing length of the
  $\pi$-conjugated system. The decrease is steeper for the local PBE
  functional in comparison to the two functionals that involve a part of
  HF-exchange. This was already noted by Gierschner et
  al.\cite{Gierschner2007} and can be traced back to the structure of
  the RPA equations in TDDFT, as we believe.  Corrections to Kohn-Sham single-particle
  excitations are brought about by the so-called coupling matrix,
  which features exchange-like two-electron integrals for pure
  functionals.\cite{Casida1995} These tend to zero for extended
  molecular orbitals, such that the excited state energies given by TDDFT
  reduce to simple molecular orbital energy differences. Within a
  particle-in-a-box model the latter scale as $1/L^2$, where $L$
  denotes the effective conjugation length. In contrast, the coupling
  matrix involves Coulomb-like two-electron integrals with $1/L$
  scaling for functionals that involve
  HF-exchange.\cite{izmaylov2008tdd}

With respect to convergence to
 the polymer limit, the T$_1$ energies of all copolymers saturate with
 some fluctuations after approximately four repeat units. Compared to
 scaling of of the range separation parameter, the
 convergence occurs faster here.   
A further observation is related to the absolute size of the triplet
excitation energies, which follow the order LC-PBE $>$ B3LYP $>$ PBE
in line with the extensive benchmark provided by Jacquemin et
al.\cite{Jacquemin2010c}. This study  indicates an underestimation of
T$_1$ energies of roughly 0.4 eV with respect to theoretical best
estimates for the present functionals. For the longest calculated
triphenylene chain {\bf P2}$_6$, the LC-PBE functional delivers a
phosphorescence energy of 2.31 eV and comes very close to the
experimental result of 2.39 eV given by Lupton. In stark disagreement
with the measurements, the different copolymers do however not
converge to the same emission energy. The large Stokes shift of
0.5 eV, on average, points towards a significant excited state
relaxation and concomitant change of excited state nature.

\subsection{\label{sec:localization}Localization behavior}

\begin{figure}[t]
        \centering
        \begin{subfigure}[t]{0.5\textwidth}
                \centering
                \includegraphics[width=\textwidth]{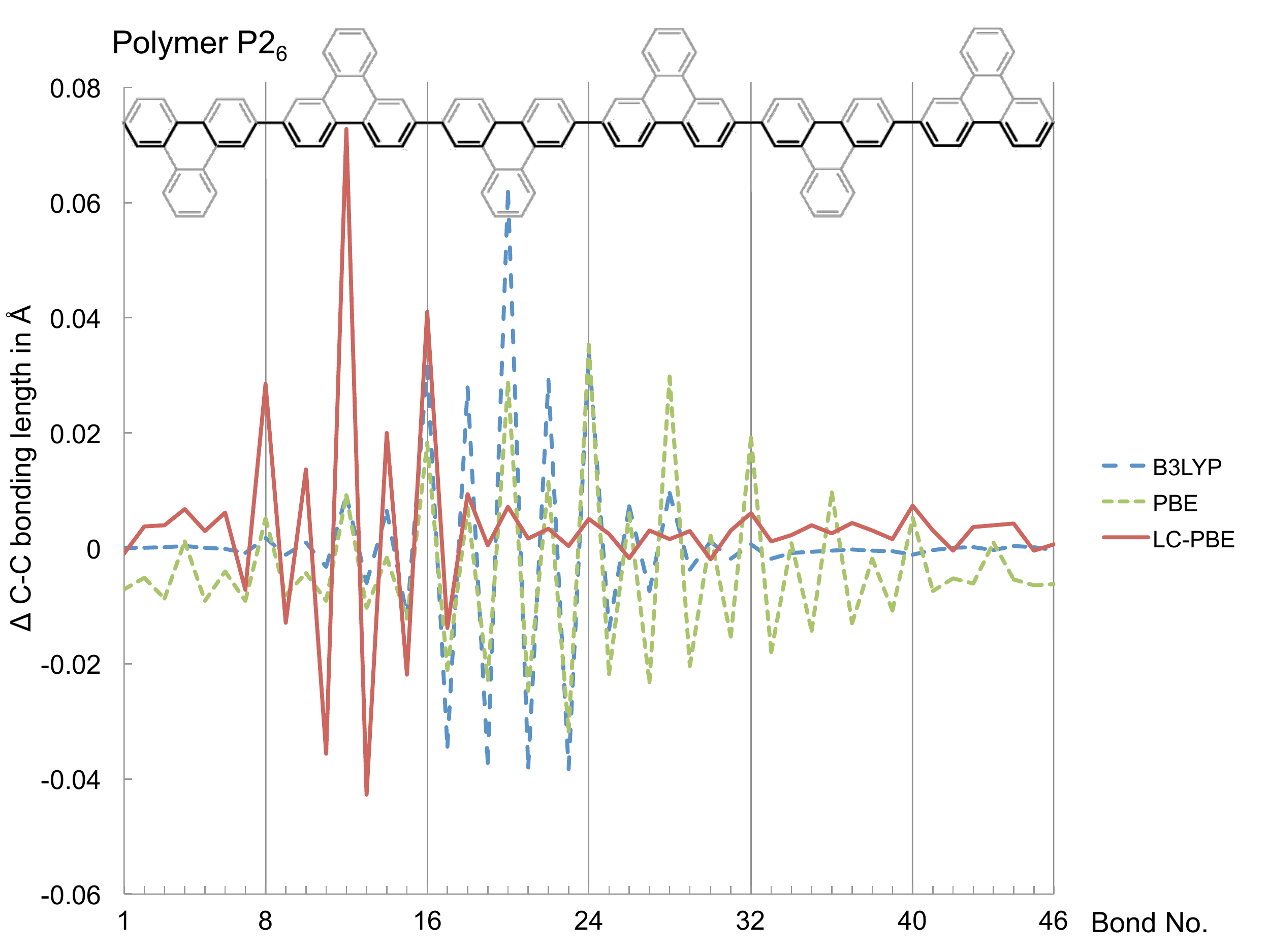}
                \caption{}
                \label{fig:geoP2}    
        \end{subfigure}%
        \begin{subfigure}[t]{0.5\textwidth}
                \centering
                \includegraphics[width=\textwidth]{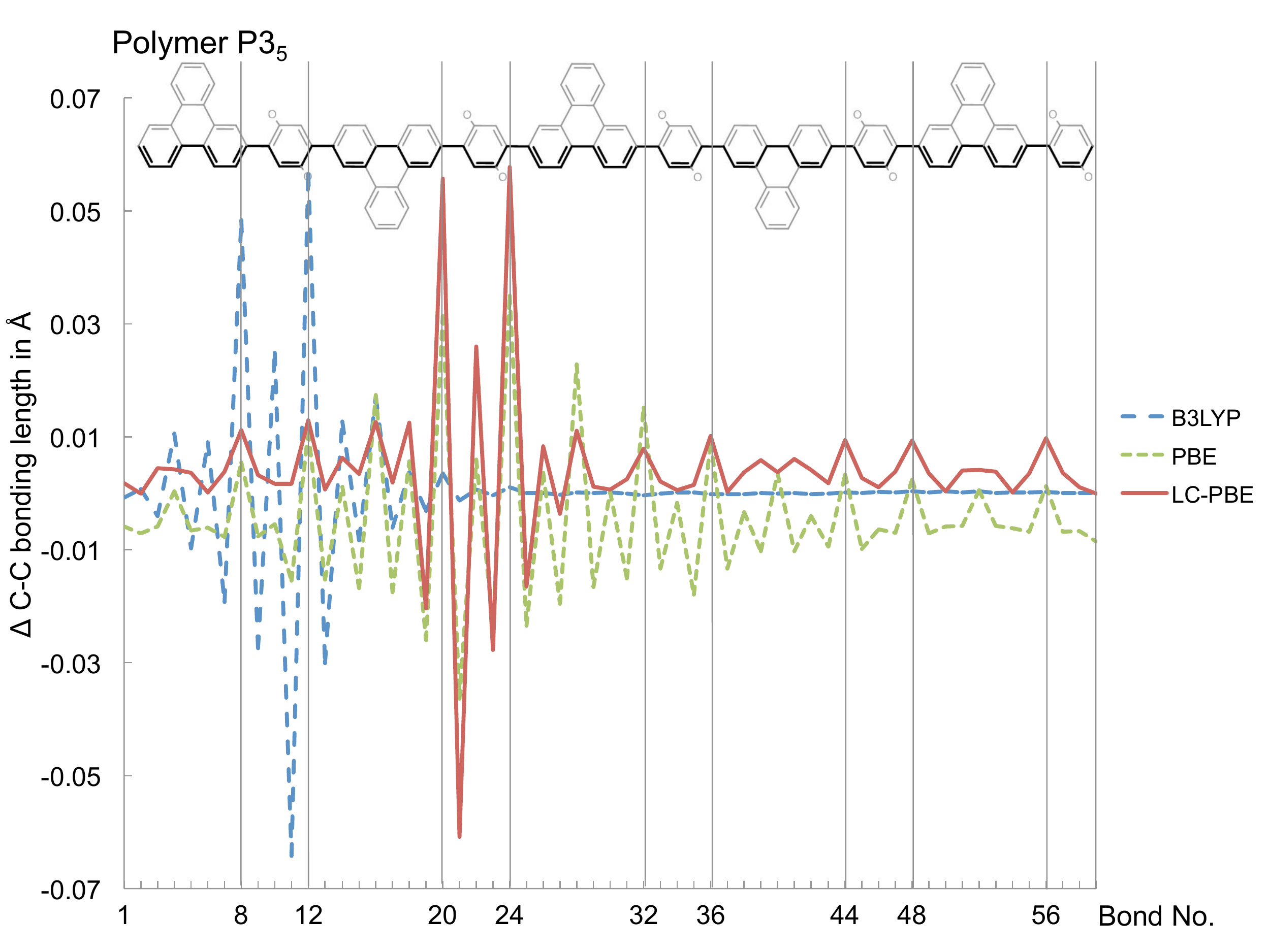}
                \caption{}
                \label{fig:geoP3}
        \end{subfigure}
        \begin{subfigure}[t]{0.5\textwidth}
                \centering
                \includegraphics[width=\textwidth]{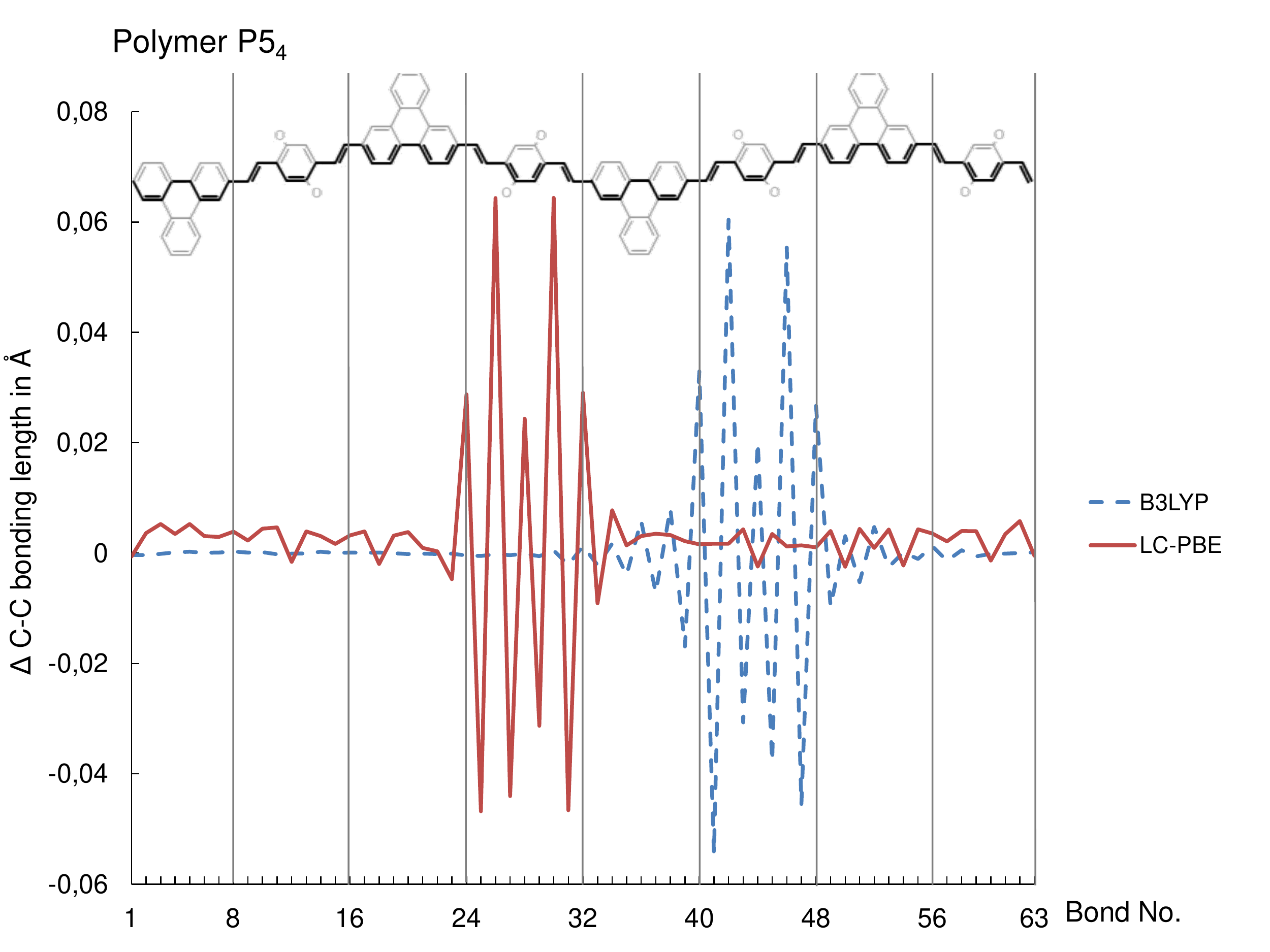}
                \caption{}
                \label{fig:geoP5}
        \end{subfigure}%
        \begin{subfigure}[t]{0.5\textwidth}
                \centering
                \includegraphics[width=\textwidth]{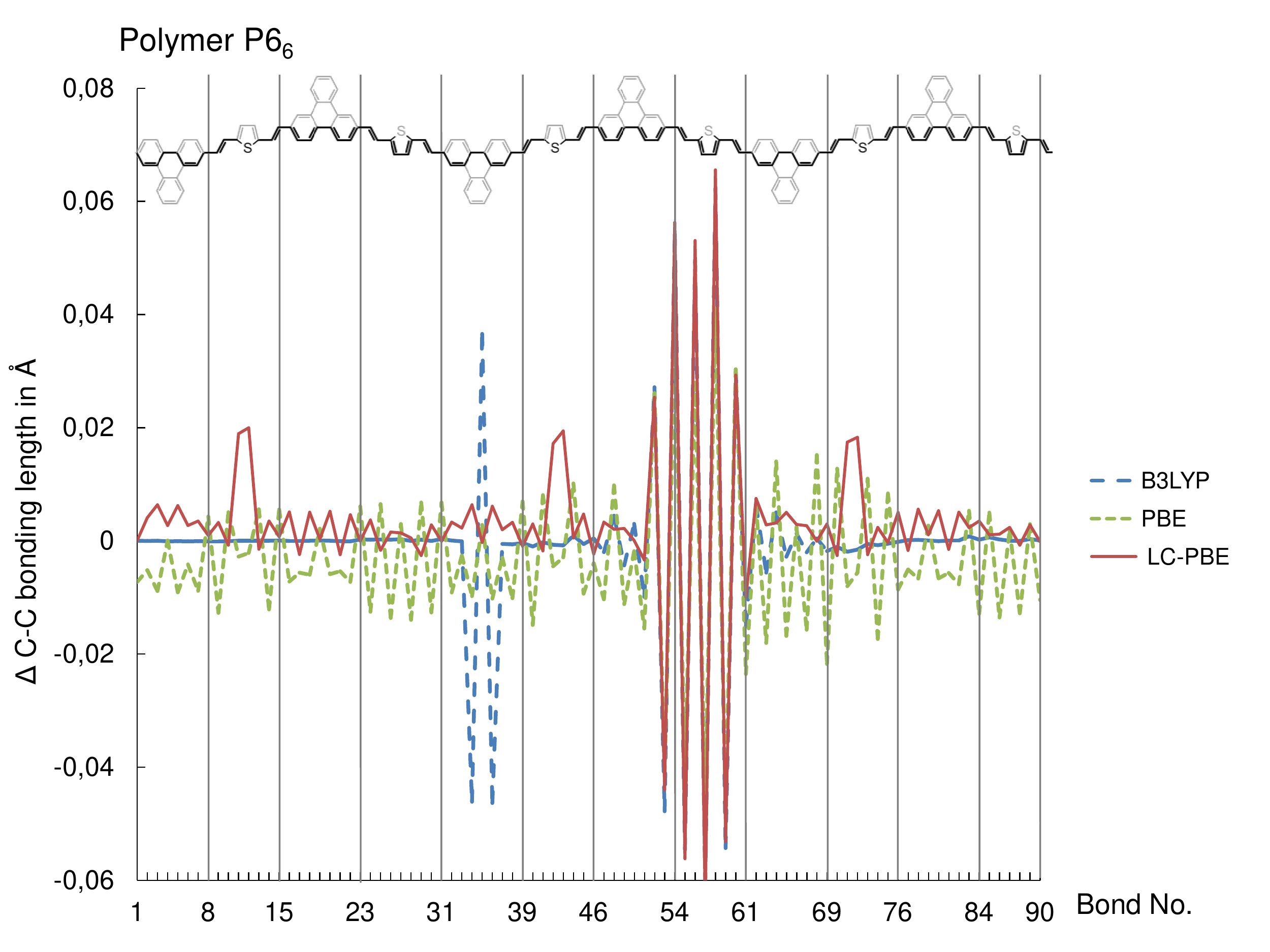}
                \caption{}
                \label{fig:geoP6}
        \end{subfigure}
\caption{C-C bond length difference between geometries optimized in the
  S$_0$ and T$_1$ state along the sketched path for the different polymers. The difference shows where bonds get tighter (positive value) or looser (negative value).}
\label{fig:geochange}       
\end{figure}

In order to quantify these geometrical changes, we follow
Ref.~[\citeonline{Pogantsch2002}] and plot the bond
length difference between S$_0$ and T$_1$  optimized geometries along the central
carbon chain in Figure \ref{fig:geochange}. Generally the bonding pattern
changes from aromatic to quinoid form. Focusing first on the pure
triphenylene polymer {\bf P2} (Figure \ref{fig:geoP2}), we find in agreement with the
experimental results, that the structural changes are
localized on one of the triphenylene units for B3LYP and LC-PBE.  In
contrast, PBE predicts bond length changes that are smeared out
over the full chain. As already mentioned in the introduction, this delocalization error can be traced back to
spurious self-interaction in the local PBE functional.

For {\bf P3} in Figure \ref{fig:geoP3}, a similar behavior is found
for the PBE functional. The hybrid B3LYP favors again a localization,
but this time not on the triphenylene moiety of the copolymer, but on
the benzoquinone. Also for {\bf P5}, both B3LYP and LC-PBE predict a
localization on the bridging units, although they differ in the actual
unit along the chain. The localization pattern for {\bf P6} follows
this trend, again the excited state is predominantly localized on the bridging unit
thiophene for B3LYP and LC-PBE, while the PBE result is fully
delocalized. Interestingly, a partial localization on a triphenylene
unit is observed for the B3LYP functional for this polymer.    

The discrepancy between theory and experiment mentioned in the last
section can now be explained. The calculations predict either a
localization on a different moiety of the polymer as the experiment
(B3LYP, LC-PBE), or no localization at all (PBE). Since the
localization occurs on different chemical units for the different
polymers, also the phosphorescence energies must vary in contrast to
the experimental results which indicate exclusively triphenylene
emission. In line with this, the LC-PBE results are in good quantitative agreement for the pure
triphenylene polymer {\bf P2} only.

\begin{figure}[t] 
\centering
\includegraphics[width=0.5\textwidth]{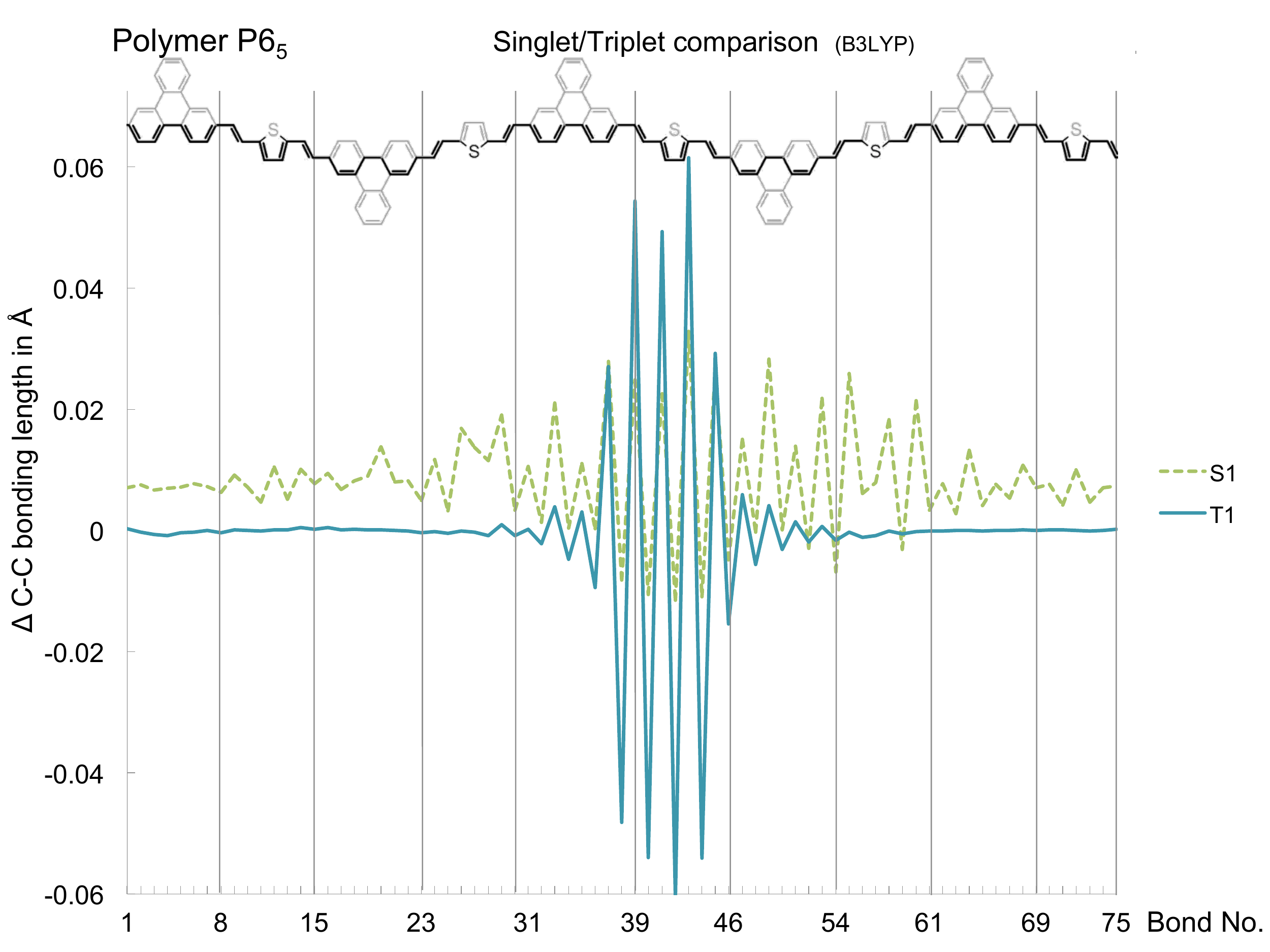}
\caption{S$_0$/S$_1$ as well as S$_0$/T$_1$ bond length difference for
  {\bf P6}$_5$ at the  B3LYP level}
\label{fig:P6_5_S1}
\end{figure}

Although the optimizations in the first singlet state could only be
performed for a subset of all polymers and functionals, we shortly discuss {\bf P6} as
an example. Figure \ref{fig:P6_5_S1} depicts a comparison of S$_1$
and T$_1$ bond length differences with respect to the ground state. It
is obvious, that the triplet excited state is more localized than the
singlet excited state, which also agrees with the experimental
findings. This can be rationalized by the fact, that the triplet state
is stabilized against the singlet due to the attractive exchange
interaction between electrons. The latter is maximized by wave function
localization, which promotes a structural change towards a confined exciton.

\section{Conclusions}
The results of the present study create mixed impressions. Some
aspects of the experiment are well reproduced, while others call for
further improvement of the computational methodology. On the positive
side, a central point of the experiment was the hypothesis that
{\em singlet and triplet excitations can form on different parts of
  the of the conjugated system}\cite{Chaudhuri2010}, which eventually
allowed the singlet-triplet gap to be tuned. The calculations confirm
this and show explicitly that the localization is not present in
absorption, but instead driven by geometrical relaxation in the
excited state. The simulations describe also the different extension
of triplet and singlet excitons qualitatively correct and
reach quantitative accuracy for the phosphorescence energy of the pure
triphenylene polymer.

On the negative side, none of the employed functionals was able to
reproduce the triplet emission at fixed energy. No clear advantage
of tuned range-separated over hybrid functionals can be ascertained in
this respect, although the absolute values for triplet emission
energies are bit more accurate for the former. It became also clear,
that local functionals like PBE are not the optimal choice for extended
$\pi$-conjugated systems. Besides their known deficiencies for these systems, like
an underestimated bond length alternation and underestimated vertical
excitation energies, we also found an  erroneous excited state
relaxation in this study. The reduced self-interaction error in hybrid and range
separated functionals helps to diminish this problem.  There is still
room for improvement, since tuned range-separated functionals violate
size consistency while hybrid functionals suffer from remaining self
interaction.\cite{Korzdorfer2012} 

Besides deficiencies in the available functionals, there are also other possible reasons for the remaining
discrepancies between theory and experiment. The first one is related
to the observation that the optimization in the triplet state locates only
the local minimum that is nearest to the Franck-Condon point. The
fact that LC-PBE and B3LYP show a localization on the same chemical
group, but on different units along the chain, hints
at several minima on the T$_1$ potential energy surface. The underlying physical picture is that of a
polymer as a multi-chromophoric system,\cite{Lupton2010} where excitons can relocate between
different chemical units by means of F\"{o}rster energy transfer. In principle,
this could be tested with excited state molecular
dynamics simulations, which are unfortunately prohibitively expensive for
the polymers under study. A related point concerns the simplified model structures we employed
in this work. In section \ref{sec:gsgeo}  we argued against a major influence of the
aromatic groups that are attached to the triphenylene
moieties. Notwithstanding, the small red-shift they induce might be
sufficient to stabilize the exciton on this moiety. Further
experimental studies on unfunctionalized triphenylene polymers could
shed further light on this question and help in the validation of
current computational approaches.  
              
\section{\label{sec:acknow}Acknowledgments}
The authors would like to thank John Lupton for fruitful discussions. Financial support from the Deutsche Forschungsgemeinschaft (GRK 1570
and SPP 1243) is also gratefully acknowledged.



\end{document}